\documentclass[oneside,11pt]{article}
\usepackage{epsfig}
\usepackage{graphicx}
\usepackage{epstopdf}
\def\ni{\noindent}
\def\om{\omega}

\def\ni{\noindent}

\def\la{\label}

\newcommand{\be}{\begin{equation}}
\newcommand{\ee}{\end{equation} \noindent}
\newcommand{\bea}{\begin{eqnarray}}
\newcommand{\eea}{\end{eqnarray} \noindent}

\usepackage{pslatex}
\usepackage{wrapfig}

\begin{document}
\bibliographystyle{unsrt}

%\title{ \vspace{1cm} Seahorse\\ \Large{A new wave energy converter}}
%\author{Francis J.M.~ Farley 
%\\
% \normalsize{8 Chemin de Saint Pierre, 06620 Le Bar sur Loup, France}}
%\maketitle
%\begin{abstract} r
%\end{abstract}
%\eject    %this makes a page break after the abstract
%\tableofcontents

%\bibliography{g2refs}
\begin{center}
%\pagestyle{myheadings}
%\markright{MED}
%I N  \hspace{3mm}     C O N F I D E N C E

% \vspace{2cm}

 {\LARGE Muons, gravity and time}
 %\vspace{1cm}
 
 Francis J.M. Farley 
 
  \emph{Energy and Climate Change Division, University of Southampton, Highfield, Southampton SO17 1BJ}
  
   F.Farley@soton.ac.uk

 \end{center}

\ni Abstract

\ni In the muon storage rings the muons are subject to a very large radial acceleration.  The equivalence principle implies a large gravity force.  It has no effect on the muon lifetime.

\ni Keywords: relativity, equivalence, gravity, time

\section{Introduction}

Does gravity affect the scale of time?  Two recent observations suggest that it does not: contradicting the established doctrine.

\ni An electron accelerated in a linac moves only a little faster, but its mass increases and therefore its energy.  Similarly a photon falling in a gravitational field picks up energy, its mass increases, its frequency rises and it becomes bluer.  Energy is conserved.

\ni With his equivalence principle Einstein \cite{E1} proposed that gravity and acceleration were indistinguishable.  He then considered light propagating vertically in an accelerating lift.  For an observer outside the lift the Doppler shifts at source and receiver are different so there is a change of frequency between the top and bottom of the lift.  An observer inside the lift is not aware of the motion or the acceleration, but thinks that he is in a gravitational field.  Attributing the change in frequency to gravity, there are two possible effects.  

\ni (a) Conservation of energy requires a falling photon to pick up energy as it drops; leading to an increase in its mass $h\nu/c^2$ and one easily derives the change in frequency $\nu$
\be
\Delta \nu / \nu = g x / c^2 = \Delta \phi / c^2
\ee
for a change in height $x$ with gravitational acceleration $g$, corresponding to a change $\Delta \phi$ in gravitational portential
 \ni The alternative explanation favoured by Einstein is
 
 \ni (b) the frequency of the falling  photon is unchanged but the scale of time is altered by gravity. Adopting this view,  Einstein derived the same formula (1).
Because of his prestige this has become the established doctrine.  Gravity changes the scale of time but has no effect on the frequency of rising or falling photons.

\ni  It is generally believed that this was confirmed by the observations of Rebka and Pound, followed by Pound and Snider \cite{RS}.  But Pound himself thought that he was measuring "the effect of gravity on gamma radiation" with no implications for the scale of time.

\ni Since then many observations have claimed to confirm the effect of gravity on time.  But all ignore the influence of gravity on the photons (messenger photons) that in most cases carry the information about frequency from one height level to another.  For example, in their experiment "Gravity Probe A" \cite{vessot}, NASA shot a hydrogen maser on a vertical ballistic trajectory and recorded the frequency of the signal received at ground level.  No correction was applied for the change of photon frequency as it travelled from the varying height of the rocket to the receiver on the ground.  Instead it was assumed that the maser frequency changed with height because the scale of time was altered by gravity.

\section{Muon storage ring}
In the muon storage rings at CERN \cite{ms1} and Brookhaven \cite{ms2} muons of $3.1\ GeV$ are trapped in a  ring magnet $14\ m$ diameter for hundreds of microseconds.  According to special relativity the  muon lifetime, normally $2.198\ \mu s$, is dilated to $64.435\  \mu s$ by the relativistic factor $\gamma = 29.327 $.  This was verified with a precision of $0.1\  \%$.  The lifetime of muons turning in the ring is equal to the lifetime of the same muons travelling in a straight line.  The transverse acceleration of the muons in orbit apparently has no effect whatsoever on the rate of muon decay!

\ni  According to the principle of equivalence the transverse acceleration should produce an effect on time due to the effective gravitational potential at the muon orbit.  The radial acceleration of the muon at orbit radius $r$ and angular frequency $\om$ is $r \om^2$.  To calculate the effective gravitational potential $V$ at the orbit assume that this gravity field, proportional to $r$ extends to the centre of the ring and integrate along a radius.  This gives
\be 
\Delta \phi = \om^2 \int r dr = (1/2) r_0^2 \om^2   \la{dv}
\ee
where $r_0$ is the radius of the storage ring.  So according to (1) we expect a slowing down of the muon clock frequency $f$ by
\be
\delta f / f = r_0^2 \om^2 / 2 c^2     \la{del1}
\ee
The muon velocity is very close to $c$, so $\om = c / r_0$ which gives
\be
\delta f / f = 0.5    \la{del2}
\ee
This implies that the muon lifetime, already dilated by special relativity from $2.2$ to $64\ \mu s$ should be lengthened by a further factor of 2. 

\ni  This estimate uses the transverse acceleration of the muon in the lab frame.  It could be more relevant to use the transverse acceleration felt by the muon in its rest frame.  The acceleration is given by the change in radial momentum of the particle divided by time.  Making a longitudinal Lorentz transformation to the muon frame the transverse momentum is unchanged, but the time is reduced by $\gamma = 29$.  Therefore in the muon rest frame its acceleration is $29$ times larger and this would lengthen the observed lifetime by a further factor of $29$.

\ni Nothing like this is observed.  So we conclude either that Einstein's principle of equivalence plays no role in the muon storage ring; or that gravity has no effect on the scale of time.

\ni In this experiment the time in the muon rest frame determines how many muons survive; it is observed directly.  No photons are involved.  The time is the time of the weak interaction. 

\section{Lithium ions stored in a magnetic ring}
The test in the Experimental Storage Ring at Darmstadt \cite{bm} was at lower $\gamma$ but time was measured much more accurately.  Metastable lithium $Li^+$  ions were stored with $\beta = 0.336$ .  Lasers were used to measure the frequencies of optical transitions and the time dilation factor $\gamma = 1.06263$ was verified with a precision of 0.0023 parts per million.  In this case eqn (3) gives
\be
\delta f / f = \beta^2/2 = 0.056
\ee
If equivalence is valid and gravity changes time, the transverse acceleration in the ring should change the optical frequencies by $5.6 \%$.  No such change was observed.  Here the time concerned is the time of quantum electrodynamics.

\section{Summary}
Notional gravity due to the radial acceleration in storage rings does not affect time.  Experiments purporting to establish the effect of gravity on time may need to be reviewed, and could include the effect of gravity on the messenger photons used to communicate frequency from one height to another.


\begin{thebibliography} {50}

\bibitem{E1} Albert Einstein, Ann. Physik, 35, 898 (1911); \emph {Relativity: The Special and General Theory}, Methuen 1924

\bibitem{RS}   Pound R. V. and  Snider J.L.  \emph{Effect of gravity on gamma radiation}, Phys. Rev. 140, B788 (1965)

\bibitem{vessot}  Vessot R. F. C.  and  Levine M. W., \emph{Gravitational redshift space probe experiment}, GP-A Project Final Report, Smithsonian Astrophysical Observatory, Cambridge, Massachusetts 02138, December 1979;  Vessot R. F. C., Levine M. W. \emph{et al.}, 
\emph{Test of Relativistic Gravitation with a Space-Borne Hydrogen Maser}
 Phys. Rev. Lett. 45,
2081 (1980)

\bibitem{ms1} Farley F.J.M. \emph{Muon g-2 and tests of relativity} , in \emph{60 years
of experiments and discoveries at CERN},  Schopper E. and  Di Lella L. \emph{eds}, World Scientific, Singapore, 2015 ;
 Bailey J. \emph{et al.} Nuclear Physics B 150, 1 (1979)
 
\bibitem{ms2} Bennett G.W. \emph{et al.} Phys. Rev. D73, 072003 (2006)

\bibitem{bm}  Botermann B. \emph{et al.}  \emph{Test of time dilation using stored $Li^+$ ions as clocks at relativistic speeds}
Phys. Rev. Lett, 113, 120405 (2014)



\end{thebibliography}
\end{document}